\documentclass[sigconf]{acmart}
\AtBeginDocument{%
  }

\usepackage{makecell}
\usepackage{subcaption}
\newcommand{\mytexttilde}{{\raise.17ex\hbox{$\scriptstyle\sim$}}}

\usepackage{enumitem}
\AtBeginDocument{%
  }

\copyrightyear{2026}
\acmYear{2026}
\setcopyright{cc}
\setcctype{by}
\acmConference[CIKM '26]{Proceedings of the 35th ACM International Conference on Information and Knowledge Management}{November 07--11, 2026}{Rome, Italy}
\acmBooktitle{Proceedings of the 35th ACM International Conference on Information and Knowledge Management (CIKM '26), November 07--11, 2026, Rome, Italy}
\acmDOI{10.1145/3799682.3840139}
\acmISBN{979-8-4007-2539-5/2026/11}

\begin{document}

\title{Mine and Refine: Optimizing Graded Relevance in E-commerce Semantic Search Retrieval}







\author{Jiaqi Xi}
\email{jiaqi.xi@doordash.com}
\affiliation{%
  \institution{DoorDash Inc.}
  \city{New York City}
  \state{New York}
  \country{USA}
}

\author{Raghav Saboo}
\email{raghav.saboo@doordash.com}
\affiliation{%
  \institution{DoorDash Inc.}
  \city{New York City}
  \state{New York}
  \country{USA}
}

\author{Luming Chen}
\email{luming.chen@doordash.com}
\author{Johny Rufus}
\email{johny.rufus@doordash.com}
\affiliation{%
  \institution{DoorDash Inc.}
  \city{Sunnyvale}
  \state{California}
  \country{USA}
}

\author{Aditya Dodda}
\email{aditya.dodda@doordash.com}
\author{Ved Sampath}
\email{ved.sampath@doordash.com}
\author{Kenny Chi}
\email{kenny.chi@doordash.com}
\affiliation{%
  \institution{DoorDash Inc.}
  \country{USA}
}

\author{Elyse Winer}
\email{elyse.winer@doordash.com}
\author{Akshad Viswanathan}
\email{akshad.viswanathan@doordash.com}
\affiliation{%
  \institution{DoorDash Inc.}
  \country{USA}
}

\author{Martin Wang}
\email{martin.wang@doordash.com}
\author{Sudeep Das}
\email{sudeep.das2@doordash.com}
\affiliation{%
  \institution{DoorDash Inc.}
  \city{San Francisco}
  \state{California}
  \country{USA}
}

\renewcommand{\shortauthors}{Xi, Saboo, et al.}


\begin{abstract}
    
    

Embedding-based retrieval (EBR) for large-scale e-commerce search faces three intertwined challenges: graded (non-binary) relevance where engagement signals are noisy and intent-varying while business relevance guidelines admit acceptable-but-not-exact matches, false negatives in hard sample mining, and unstable similarity score separability across relevance levels, the last of which complicates hybrid search score fusion and downstream ranking. 
We propose \emph{Mine and Refine}, a two-stage contrastive training framework that addresses all three. A lightweight LLM, fine-tuned with engagement-driven audit, serves as a guideline-aligned scalable labeler throughout training. 
Stage 1 establishes a robust global embedding space via label-aware supervised contrastive learning; Stage 2 mines hard samples, re-annotates them with the LLM labeler to mitigate spurious negatives, and refines the model through a multi-level extension of circle loss that enforces margin-controlled separation across relevance levels. 
Deployed in production e-commerce search across multiple product verticals, the approach delivers statistically significant lifts in user engagement and gross order value, and substantially improves retrieval and end-to-end relevance metrics.
\end{abstract}





\maketitle

\section{Introduction} 
\label{sec:intro}
Semantic retrieval has become a core component of modern e-commerce search, enabling systems to retrieve relevant items even when user queries do not lexically match product text. 
In a multi-category marketplace, this problem is compounded by heterogeneous catalog quality, rapidly changing inventory, and a query distribution dominated by long-tail and noisy inputs (misspellings, abbreviations, multilingual queries), all under strict latency and indexing constraints typical of production retrieval. 

Despite strong progress in embedding-based retrieval (EBR), deploying a robust semantic retriever in e-commerce presents three practical challenges.

\emph{First, relevance is graded and governed by domain specific guidelines.}
Shopping queries admit exact matches, substitutes, complements, and irrelevant results, requiring labels that align with business definitions of relevance rather than engagement signals alone.

\emph{Second, scalable training requires hard mining, but naive mining introduces noise.}
Top ANN candidates right below positive samples are often \emph{near misses} that remain relevant in e-commerce context (e.g., substitutes or complements). 
Treating them as plain negatives corrupts training signals and harms generalization. 
At the same time, high-quality human labels at scale are costly.

\emph{Third, intra-level compactness and consistent inter-level separability of similarity scores are essential for production stability.}
Embedding similarity scores are fed into hybrid score fusion logic, and the query and item embeddings themselves serve as features in downstream ranking models. 
Well-defined boundaries between relevance levels and query-stable score distributions reduce signal noise for both, improving robustness under noisy candidate sets.

To address these, we propose a two-stage \emph{Mine and Refine} fine-tuning framework under graded relevance schema. 
We fine-tune a lightweight LLM on human-annotated data with engagement-guided auditing to produce a scalable, guideline-aligned labeler.
In \textbf{Stage~1}, a two-tower retriever is fine-tuned with a label aware supervised contrastive objective to shape a robust global semantic space. 
In \textbf{Stage~2}, hard samples are mined via ANN search, re-annotated by the LLM to keep signal consistency, and the model is further refined with a \emph{multi-level extension of circle loss} that sharpens similarity boundaries between relevance strata. 

We evaluate the proposed approach using both offline relevance metrics and production A/B experiments. 
Offline, our method improves NDCG, recall, and precision across multiple retrieval depths, with better score separability in high lexical-overlap cases.
Online A/B tests yield statistically significant lifts in add-to-cart rate, conversion rate, and gross order value when swapping only the embedding retriever while keeping lexical retrieval, downstream ranking architecture, and business logic unchanged.

\textbf{Contributions.} This paper makes the following applied contributions: 
(i) an engagement-audited efficient LLM labeler for graded relevance in line with business relevance guidelines; (ii) a two-stage \emph{Mine-and-Refine} fine-tuning framework with LLM-annotated hard mined samples; (iii) a multi-level extension of circle loss enforcing per-level margins; (iv) large-scale production validation with statistically significant gains.

Along the way, several counterintuitive findings emerge with broader applicability. 
First, regularizing embeddings between clean queries and their misspelled variants, which seems to be a natural choice for invariance, degrades robustness rather than improves it, while additive augmentation that preserves both forms works substantially better. 
Second, randomly injecting synthetic queries improves retrieval metrics but degrades relevance score separability. 
This tradeoff is invisible to standard evaluation, but consequential for production hybrid retrieval.

The remainder of the paper is organized as follows. Section~\ref{sec:related-work} reviews related work. Section~\ref{sec:label} describes our labeling framework. Section~\ref{sec:model} details model architecture, training data construction, and training techniques. Section~\ref{sec:experiments} presents the offline and online evaluation results, followed by ablation studies and discussion. Section~\ref{sec:parameters} presents training details including hyper-parameter choices.

\section{Related Work}
\label{sec:related-work}

\noindent \textbf{Embedding-based Retrieval.}
Lexical methods such as BM25 \cite{Robertson09} remain strong baselines but struggle with semantic mismatch in e-commerce search. 
Embedding-based retrieval (EBR) addresses this by encoding queries and items into a shared vector space using bi-encoder architectures and retrieving candidates via ANN search \cite{Reimers19, Karpukhin20}. 
EBR is widely adopted for its latency–quality trade-off and has been extended with richer interaction mechanisms while preserving offline indexing \cite{Khattab20}. 
Multilingual support is typically achieved via multilingual encoders that align languages in a shared semantic space \cite{Yang19, Feng20, Wang22}.

\noindent \textbf{Contrastive Learning.}
Contrastive learning objectives are central to learning retrieval embeddings, from early triplet and N-pair losses \cite{Weinberger09a, Sohn16} to modern InfoNCE-style objectives that contrast one positive against many in-batch negatives \cite{Gao21, Izacard21}. 
Supervised contrastive learning (SupCon) \cite{Khosla20} generalizes this to multi-positive settings with improved stability. 
Separately, margin-based objectives such as ArcFace \cite{Deng19} explicitly encourage separation in similarity space and circle loss \cite{Sun20} provides a unified similarity optimization framework through adaptive margin control.
However, these approaches typically assume binary relevance. Our framework incorporates three-level graded relevance directly into the objective: a multi-level supervised contrastive stage learns global structure, followed by a circle-style refinement adapted to graded labels to enforce sharper similarity boundaries.

\noindent \textbf{Hard Sample Mining and Iterative Refinement.} 
Negative sample quality is critical for EBR. 
In-batch negatives are efficient but prone to false negatives, especially under broad intents or graded relevance. 
Stronger negatives are obtained via offline or iterative mining, where a retriever selects hard candidates for further training~\cite{Xiong21}; broader iterative refinement frameworks alternate between retrieval and retraining on mined candidates, often combined with curriculum learning~\cite{Karpukhin20, Gao21, Izacard21, Xiong21, Wang22}. 
However, aggressive mining can amplify label noise, and prior pipelines largely assume binary relevance. Our framework departs in two ways: (i) an explicit LLM-based annotation step labels newly mined pairs, enabling reliable selection of both hard negatives and hard positives in a three-level setting; and (ii) refinement applies a geometry-aware objective, multi-level circle loss to sharpen inter-class similarity margins, targeting robust score calibration under graded relevance, which is critical for stable hybrid fusion and downstream ranking, yet largely unaddressed by prior binary-relevance pipelines.

\section{Labeling Framework} \label{sec:label}
Embedding-based retrieval in e-commerce has been deployed at multiple major platforms, each with distinct design choices for handling label quality and relevance signals. 
Most approaches treat engagement signals as the primary supervision, augmenting them in various ways: heuristic mining guardrails such as product-type filtering or query-token overlap~\cite{Huang20, Li21}, a cross-encoder relevance reward model trained on human annotations and incorporated through label revision and a multi-objective loss~\cite{Lin24}, and category- or attribute-based augmentation~\cite{Nigam19, Xie22}. 

However, engagement signals vary with query breadth, broad queries exhibit lower conversion rates due to dispersed user intent, making universal thresholds unreliable.
They are also inherently noisy, reflecting not only relevance but factors such as presentation, promotion, and user preference. 
More systematic methods \cite{Yao21} partially address these issues but remain less accurate and efficient than lightweight LLM fine-tuning.

Recent works~\cite{mehrdad2024, faggioli2023} have explored several techniques for leveraging LLMs to automate query-item relevance judgment in product search. 
Our labeling framework builds on this direction with two extensions: (1) an engagement-driven audit stage that filters human label noise before LLM fine-tuning, addressing both the breadth-dependent reliability of engagement thresholds and the conflation of relevance with other factors; and (2) integration of the LLM labeler into a full embedding retriever training pipeline as guideline-consistent supervision for both stages.

Given a query $q$ and an item $d$, where each item $d$ has attributes such as name, category path, description, etc., our relevance guidelines assign each $(q, d)$ pair one of three labels $2$ (relevant), $1$ (moderately relevant -- substitute or complement), and $0$ (irrelevant). 

\subsection{Engagement Audit} \label{sec:audit}
We construct the following LLM labeler fine-tuning design.
Similar to the Amazon shopping data set \cite{Amazon22}, we begin with $\mytexttilde 705$k human annotated query--item pairs under this three-level schema as ground truth, and then audit them against engagement signals by sampling $\mytexttilde30\%$ of cases with label–behavior disagreement (e.g., low relevance but high conversion, or vice versa).
These cases are re-evaluated using stronger LLMs (e.g., GPT-4o \cite{OpenAI24}, o3 \cite{OpenAI25}, Gemini-2.5-Flash \cite{Gemini25}) with more granular and structured prompts. 
Conflicts between human and LLM audit labels are resolved using a Query-to-Category (Q2C) transformer model that predicts relevant categories from query intent: if the item's category falls within the Q2C predictions, we take the higher of the two labels; otherwise, the lower. 
Under this resolution, the audit LLM agrees with the final label in $81.8\%$ of conflict cases, and the one-time correction reduces label errors by $5.74\%$.

\subsection{Fine-tuned LLM for Scalable Labeling}
A held-out evaluation set of $105$k query--item pairs is sampled from the audited data in a stratified way spanning diverse categories and intent breadth.
We fine-tune GPT-4o mini~\cite{OpenAI24} on the remaining $\mytexttilde 600$k pairs.
The fine-tuned LLM achieves $87.6\%$ exact accuracy and $98.8\%$ within-1 accuracy (predictions off by at most one level) against audited human annotations in the evaluation set. 
This enables scalable relevance labeling for over $100$M query–item pairs used in embedding model training.

\section{Semantic Embedding Model} \label{sec:model}


\subsection{Model Architecture}

\begin{figure*}[htbp]
  \includegraphics[width=\textwidth]{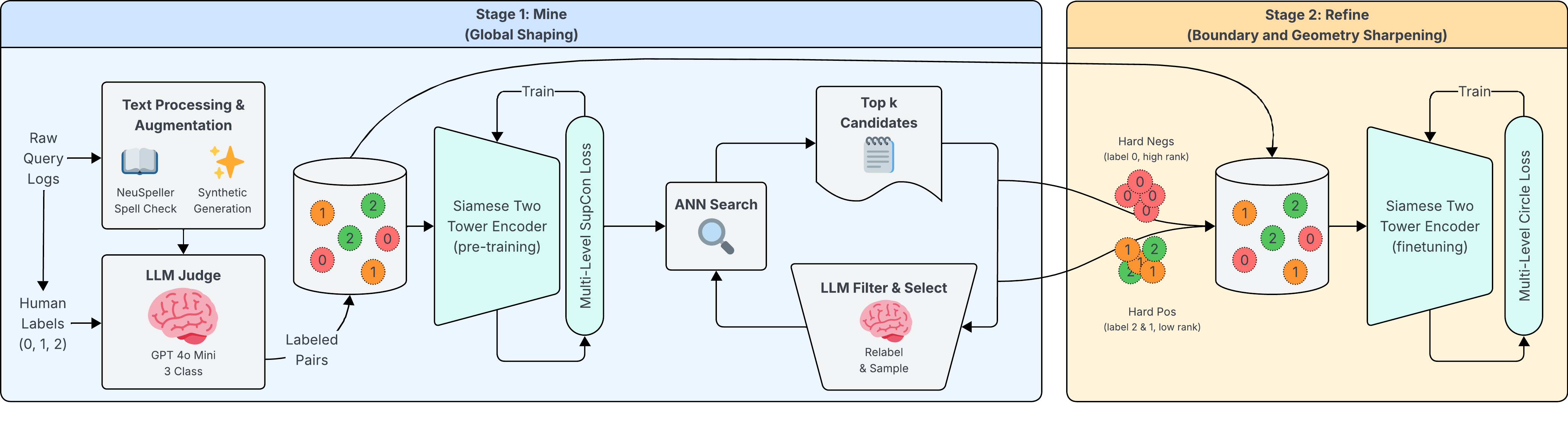}
  \caption{Mine and Refine Overview: Stage 1 trains a shared encoder on the original dataset using supervised contrastive (SupCon) loss. Stage 2 continues training on a curriculum dataset using circle loss with self-pacing sample weighting and a definite convergence target induced by circular decision boundary.}
  \Description{}
  \label{fig:train-overview}
\end{figure*}

As shown in Figure~\ref{fig:train-overview}, we adopt a variant of the two-tower architecture, specifically a Siamese encoder with two separate projection heads that compress embeddings into a lower ($256$) dimensional subspace. 
We choose the Siamese encoder due to its demonstrated advantages over asymmetric encoders \cite{Dong22} and its widespread adoption in open source pretrained models (e.g., E5~\cite{Wang2024}, BGE~\cite{bge2024}, and GTE~\cite{mgte2024}), which is also validated in our study (Table \ref{tab:structure-catalog-eval}).
The projection heads are introduced to enable efficient real-time inference under practical infrastructure and latency constraints.

The Siamese encoder is initialized from the pretrained $0.1$B multilingual-E5-small~\cite{Wang2024} model. 
Although our optimization in this paper aims at overall retrieval quality, the multilingual architecture provides a natural foundation for extending to non-English and cross-lingual retrieval scenarios in future improvements. 

\subsection{Training Data} \label{sec:data}

\subsubsection{Enrichment with Catalog Attributes}
Item representations are constructed from the concatenation of item name and a two-level category path. 
This helps the model learn a consistent pattern and thus better identify the item attributes.

Approaches to improving query understanding include attribute-level embedding composition on both query and item sides \cite{Huang20, Chen25} and synthetic query generation from item features \cite{Xie22}. 
However, embedding composition is unsuitable for real-time retrieval due to the latency of sequential attribute extraction and encoding.

In contrast, synthetic queries are well-suited to our setting. 
In addition to generating synthetic queries from item attributes and pair each query with its source item to create candidate positive pairs, we construct potential negative pairs by matching items with queries generated from different items. All synthetic query--item pairs are then labeled using the fine-tuned LLM.

\subsubsection{Spelling Variation Augmentation}
Misspelled queries are common in e-commerce search. 
While dedicated spell-correction pipelines address most cases, residual errors still degrade semantic understanding. 
To improve robustness, we augment training data with spelling variations. 
Prior approaches include replacing a subset of clean queries with typo-injected variants \cite{Zhuang21, Lin24} or enforcing invariance via cosine similarity regularization between clean and perturbed queries.

We instead develop an additive augmentation strategy that preserves the original instances while introducing misspelled variants. 
As shown in Table~\ref{tab:spell-eval}, this approach, as detailed below, achieves comparable performance on misspelled query understanding and preserves global semantic structure:
\begin{itemize}
\item Generate spelling variations for all training queries using the probabilistic character replacement noiser from \textsc{NeuSpell} \cite{neuspell20};
\item Randomly sample one variation per query, retaining $30\%$–$50\%$ of the generated variants;
\item Construct additional training instances by duplicating the original query–item pairs and assigning the sampled variation as the anchor.
\end{itemize}

\subsection{SupCon for Graded Relevance} \label{secl:supcon}
The training dataset comprises instances constructed from LLM-labeled query--item pairs, in one of the following formats
{
\setlength{\abovedisplayskip}{3pt}
\setlength{\belowdisplayskip}{3pt}
\begin{equation}\label{eq:tuples}
    (q, d^{(2)}, d^{(1)}, d^{(0)}),\,
    (q, d^{(2)}, d^{(0)}),\,
    (q, d^{(1)}, d^{(0)}),\,
    (q, d^{(2)}, d^{(1)})
\end{equation}
}
where $d^{(\ell)}$ denotes items with relevance label $\ell \in \{0,1,2\}$ corresponding to the associated query $q$.
Specifically, for each query, positive items (label 2 and/or 1) are partitioned into small balanced groups, each paired with a larger number of negatives randomly sampled from strictly lower relevance levels to form a training instance. 
Each query therefore yields multiple instances, and instance lengths vary across the dataset.

Given the nature of 3-level labels, regular binary loss functions (e.g., binary cross entropy loss, InfoNCE) do not apply directly.
In addition, the structure of many positives and many negatives for each anchor $q$ in one instance renders the triplet \cite{Weinberger09a} and N-pair \cite{Sohn16} losses unsuitable.

The supervised contrastive loss \cite{Khosla20} is employed due to its compatibility with the training data and more importantly its stability on hard negative samples and robustness to choices of hyper parameters.
While initially defined for binary labels, the supervised contrastive loss can be naturally extended to multi-level labels: 

For each training instance that anchors a query $q$ with embedding $e_q$, let $P(q)$ denote the set of items with label $2$ or $1$, $e_{d_i}$ the normalized embedding of the $i$-th item $d_i \in P(q)$, and $r_i$ its corresponding label value.
Let $e_{d_1},...,e_{d_{N_q}}$ denote the set of all item embeddings in this instance.
The loss corresponding to this instance is then defined as
{
\setlength{\abovedisplayskip}{3pt}
\setlength{\belowdisplayskip}{3pt}
\begin{equation}
    \mathcal{L}_{q} = - \frac{1}{\sum \limits_{i \in P(q)} r_i} \sum \limits_{i \in P(q)} r_i \log\Big(\frac{\exp(\langle e_q, e_{d_i} \rangle/\tau)}{\sum_{j=1}^{N_q} \exp(\langle e_q, e_{d_j}\rangle/\tau)} \Big),
\end{equation}
}
where $\tau$ denotes the temperature factor trained along with the model to allow more flexibility and smoother convergence.

Additionally, we observe consistent findings, in line with~\cite{Liu21} that $\textit{reduction} = \textit{sum}$ has more stable convergence and better performance than $\textit{reduction} = \textit{mean}$ and our conjecture is that $\textit{reduction} = \textit{sum}$ captures the varying instance sizes across batches.

\subsection{Hard Sample Mining and Margin-based Refinement} \label{sec:circle}
\subsubsection{Offline Mining Strategies.} Mining hard negatives is critical for improving retrieval quality because it surfaces model weaknesses in regions underrepresented by current training data~\cite{Huang20, Liu21, Lin24}. 

Despite the wide adoption of in-batch negatives, this strategy implicitly assumes that an item relevant to one query is irrelevant to other randomly sampled queries, which is fragile under a three-level relevance schema where cross-sample relevance is ambiguous. 
We therefore rely on offline global mining instead, trading some efficiency for cleaner and more consistent supervision.

Our proposed offline mining strategy is presented as follows:
\begin{itemize} 
    \item Perform ANN search over the training index using the Stage 1 model and retrieve the top-$k$ items per query; $k \in [100, 200]$ empirically balances efficiency and quality.
    \item Label all retrieved $(q,d)$ pairs not in the original training set with the fine-tuned LLM (Section~\ref{sec:label}) to avoid false negatives common in ANCE-style mining~\cite{Xiong21, Lin24}.
    \item From the top-$k$ results, select (i) \emph{hard negatives}: $(q,d)$ pairs with label $0$ ranked above $k/2$, and (ii) \emph{hard positives}: $(q,d)$ pairs with label $2$ or $1$ ranked below $k/2$. 
    \item Augment the original training data with the mined hard samples, and construct Stage 2 training instances from this expanded set.
\end{itemize}
While prior work is not always explicit on whether to retain or replace original training negatives during curriculum refinement~\cite{Liu21, Lin24}, we find that retaining them substantially improves performance. 
This preserves the diversity of the original data and helps prevent catastrophic forgetting.
The variant $\text{(S, SC)} + \text{(A, TL)}^H$ in Table \ref{tab:structure-catalog-eval} shows the decline in the case when negatives from the original data are discarded during curriculum training.

Furthermore, offline mining with LLM-annotation naturally introduces new positive samples, in addition to semi-positives and negatives, enriching the embedding space with more diverse query--item alignments.

\subsubsection{Dynamic Optimization and Margin Sharpness: Circle Loss.} 
Various loss functions have been explored for curriculum training \cite{Liu21, Xie22, Lin24}, including cross-entropy, softmax-based objectives, and triplet loss. 
However, none consistently achieves both stable optimization and semantically coherent embeddings with strong intra-class compactness. 
In particular, separability of cosine similarity across the three relevance levels, independent of queries, is critical for online serving, where clear score margins simplify hybrid search score fusion and ranking decisions. 

Motivated by advances in deep metric learning and face recognition, such as Angular Margin Softmax~\cite{Liu17}, Additive Margin Softmax~\cite{Wang18}, ArcFace~\cite{Deng19}, and circle loss~\cite{Sun20}, we adopt circle loss as it provides a unified objective that bridges contrastive and classification-style losses. 
By jointly modeling positive and negative similarities with adaptive weighting, it improves inter-class separability while maintaining stable training convergence.

However, the standard instantiation of circle loss assumes binary supervision and does not directly accommodate a three-level relevance schema. 
We therefore extend it to a multi-level formulation, enabling it to better align with graded relevance signals.

For any training instance $g$, it falls in one of the last three cases in equation \eqref{eq:tuples}, i.e. $g \in (q, d^{(k)}, d^{(\ell)})$ for some $k > \ell,\, k, \ell \in \{0, 1, 2\}$.
Let $\mathcal{I}_g$ denote the set of indices $j$ of the pairs $(q,d_j)$ belonging to the instance $g$, $\mathcal{I}_{g, \lambda} := \{j \in \mathcal{I}_g:\, r_j = \lambda\},\, \lambda \in \{0, 1, 2\}$, and $s_j := \langle e_q, e_{d_j} \rangle$. 
The circle loss of $g$ is defined as
{
\setlength{\abovedisplayskip}{2pt}
\setlength{\belowdisplayskip}{2pt}
\begin{align} \label{eq:circle}
        \mathcal{L}_{k, \ell}  & = \log \Bigl[ 1 + \frac{1}{\left| \mathcal{I}_{g,k}\right|}\sum_{i \in \mathcal{I}_{g, k}} \exp\Bigr(-\gamma \max(O_{k,p} - s_i, 0)(s_i - \Delta_{k, p})\Bigr)  \nonumber \\
        &  \qquad \qquad \qquad \quad \sum_{j \in \mathcal{I}_{g, \ell}}\exp\Bigr( - \gamma \min(O_{\ell, n} - s_j, 0)(s_j - \Delta_{\ell, n})\Bigr)\Bigr].
    \end{align} 
}
Then the loss of a batch is aggregated by $\textit{reduction} = \textit{sum}$.

\begin{figure}[!t]
  \centering
  \begin{subfigure}{0.49\columnwidth}
    \centering
    \includegraphics[width=1.1\linewidth]{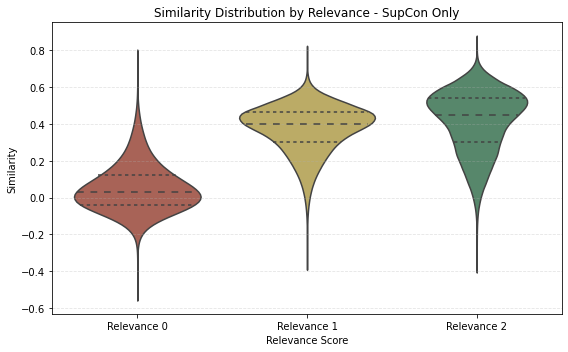}
    \caption{SupCon only}
  \end{subfigure}\hfill
  \begin{subfigure}{0.49\columnwidth}
    \centering
    \includegraphics[width=1.1\linewidth]{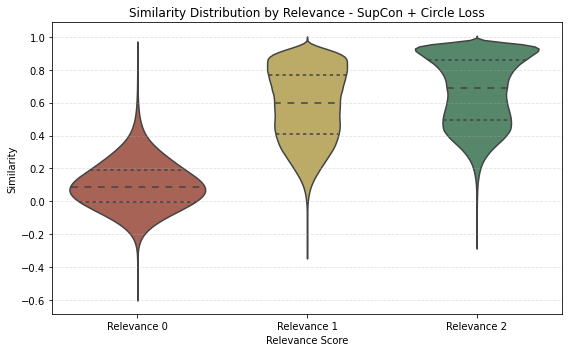}
    \caption{SupCon $+$ Circle Loss}
  \end{subfigure}

  \vspace{0.5em}

  \begin{subfigure}{\columnwidth}
    \centering
    \includegraphics[width=0.83\linewidth]{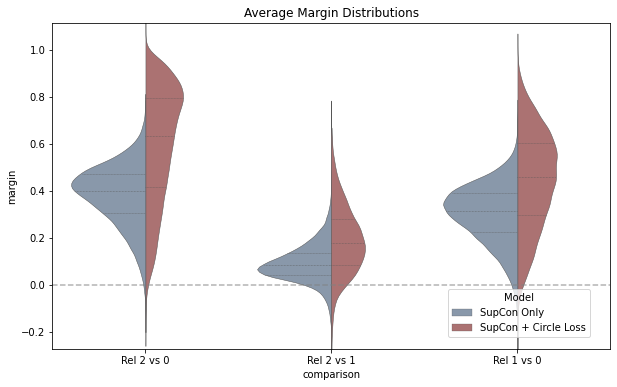}
    \caption{Query-level average margin distributions}
  \end{subfigure}

  \caption{Score separability improvement from Stage~1 to Stage~2. \\
    Top row: Violin plots of similarity score distributions by relevance level. \\
    Bottom row: Distributions of query-level average margin:
    $\mathbb{E}_{d^{(2)}}\langle q,d^{(j)}\rangle
     - \mathbb{E}_{d^{(0)}}\langle q,d^{(i)}\rangle,\,
     i > j,\, i,j \in \{0,1,2\}$.}
  \Description{}
  \label{fig:separability}
\end{figure}

Per-level margin targets $\Delta_{k,p}$ (resp. $\Delta_{k,n}$) define explicit lower (resp. upper) convergence boundaries for each relevance level $k$. 
The loss encourages positive similarities to exceed $\Delta_{k,p}$ and negative similarities to remain below $\Delta_{k,n}$. 
The scale factor $\gamma$ controls overall gradient magnitude, while $O_{k,p}$ and $O_{k,n}$ modulate per-sample contribution: samples far from their target similarity incur larger penalties, whereas samples already near the target contribute less. 
This adaptive weighting concentrates gradients on boundary-violating pairs, leading to tighter intra-level clustering and larger inter-level separation (Figure~\ref{fig:separability}). 
Following~\cite[Section~4.6]{Sun20}, all hyper-parameters are set deterministically to ensure predictable convergence and consistent similarity distributions. 
Specific values are reported in Section~\ref{sec:parameters}.

\textbf{Why two stages are complementary.} SupCon in Stage~1 establishes a globally coherent embedding space but does not enforce absolute similarity boundaries.
Circle loss in Stage~2 addresses this directly (Figure~\ref{fig:separability}).
Quantitatively, the average similarity gap between label~2 and 1 increases from 0.10 to 0.18 after Stage~2, while the average gap between label~1 and~0 increases from 0.33 to 0.44, confirming that circle loss improves separation where adjacent levels are hardest to distinguish. 
We report raw similarity scores without per-query normalization, as circle loss improves not only within-query separation but also cross-query score consistency, which is critical for stable score fusion in production.

We compare circle loss against triplet loss rather than SupCon in Stage 2 because the refinement stage requires explicit control over inter-class separation targets, a property provided by circle loss's deterministic margins and shared to a lesser degree by triplet loss's fixed margin, but absent in SupCon's relative similarity optimization.
That replacing circle loss with triplet loss degrades performance (Table~\ref{tab:structure-catalog-eval}) confirms that the multi-level margin formulation, not curriculum training alone, drives the improvement.

\section{Evaluation and Experiments} \label{sec:experiments}

We evaluate the proposed \emph{Mine and Refine} embedding retriever through a combination of offline relevance metrics and online A/B experiments.
Due to confidentiality constraints, all metrics in Tables 1–5 are reported as relative improvements over the respective baselines, and absolute values are omitted.

\subsection{Experimental Setup}
\label{sec:exp-setup}

\textbf{Task and serving setting.}
Offline studies evaluate the embedding model in isolation under controlled settings. 
We report Precision@$K$ and Recall@$K$ (along with NDCG@$K$ for selecting the final production candidate) for $K\in\{10,50,100\}$.

The production system integrates the semantic search embedding model as a complementary layer alongside a fixed lexical retrieval component.
We maintain a pre-computed HNSW-based index covering over $300$M items across the full catalog.
At request time, query embeddings are generated on the fly and matched against this index via ANN search.
We report NDCG@$10$ for end-to-end production relevance performance, as well as engagement and business metrics from online A/B experiments.
All online experiments are conducted under the same latency and capacity constraints as the baseline system.

We compare the following three retrieval systems in production:
\begin{itemize}
    \item \textbf{Lexical baseline:} a token-based retriever (BM25 style) that is kept unchanged across all three retrieval systems.
    \item \textbf{Hybrid baseline:} a hybrid search setup combining the lexical baseline with a text embedding model (the baseline $\text{(A, TL)}$ in Table \ref{tab:structure-catalog-eval}) of the same dimensionality as the proposed model, initialized from the same pretrained backbone encoder and fine-tuned using a weighted triplet loss adapted from~\cite{Huang20}, with minor modifications for three-level relevance labels. 
    \item \textbf{Hybrid w/ Mine and Refine:} the hybrid search setup combining the lexical baseline and the proposed embedding model optimized by the \emph{Mine and Refine} approach.
\end{itemize}

\begin{table*}[t]
\centering
\caption{Offline Model Evaluation across Metrics}
\label{tab:offline-main}
{\small 
\begin{tabular}{l|l|l|ccc|ccc|ccc}
\toprule
& & &
\multicolumn{3}{c|}{Recall@K}
& \multicolumn{3}{c|}{Precision@K}
& \multicolumn{3}{c}{NDCG@K} \\
\cmidrule(lr){4-6}
\cmidrule(lr){7-9}
\cmidrule(lr){10-12}
Model
& Size
& Dim
& $10$ & $50$ & $100$
& $10$ & $50$ & $100$
& $10$ & $50$ & $100$ \\
\midrule

{\small multilingual-E5-small~\cite{Wang2024}} 
& 0.1B & 384 
& - & - & -
& - & - & -
& - & - & - \\

{\small GTE-multilingual-base dense~\cite{mgte2024}} 
& 0.3B & 768 
& 4.16\%  & 5.51\% & 5.35\%
& 4.21\% & 5.88\% & 6.23\%
& 4.06\%  & 5.43\% & 5.68\% \\

{\small multilingual-E5-large~\cite{Wang2024}} 
& 0.6B & 1024 
& 4.10\% & 4.78\% & 4.58\%
& 4.11\% & 4.90\% & 5.01\%
& 4.13\% & 4.97\% & 5.09\% \\

{\small BGE-m3 dense~\cite{bge2024}} 
& 0.6B & 1024 
& -0.86\% & 0.98\% & 1.37\%
& -0.96\% & 0.88\% & 1.50\%
& -1.10\% & 0.52\% & 1.02\% \\

{\small Qwen3-Embedding-0.6B~\cite{qwen2025}} 
& 0.6B & 1024 
& 3.28\% & 5.44\% & 5.46\%
& 3.36\% & 5.71\% & 6.23\%
& 3.17\% & 5.17\% & 5.59\% \\

{\small (Siamese, SupCon)} 
& 0.1B & 256 
& 7.39\% & 14.15\% & 14.78\%
& 8.26\% & 16.11\% & 17.70\%
& 7.53\% & 13.82\% & 15.25\% \\

{\small (Siamese, SupCon) + (Asymmetric, circle loss)} 
& 0.1B & 256 
& 9.55\% & 15.87\% & 16.02\%
& 10.54\% & 18.20\% & 19.52\%
& 9.71\% & 15.91\% & 17.10\% \\

\textbf{{\small (Siamese, SupCon) + (Siamese, circle loss)}} 
& 0.1B & 256 
& \textbf{10.08\%} & \textbf{16.19\%} & \textbf{16.17\%}
& \textbf{10.96\%} & \textbf{18.46\%} & \textbf{19.69\%}
& \textbf{10.39\%} & \textbf{16.41\%} & \textbf{17.50\%} \\

\bottomrule
\end{tabular}
}
\end{table*}

\subsection{Relevance Evaluation} \label{sec:offline-eval}
We show that the proposed model achieves substantially better performance, both standalone compared to strong pretrained models and incremental impact on overall search relevance when integrated into the end-to-end system. 

\subsubsection{Golden Eval Set: Comprehensive Query Set with Small Index} 
Offline evaluation is conducted on a held-out set of $155$M query--item pairs labeled by the fine-tuned LLM (Section~\ref{sec:label}).
To better reflect real-world search behavior, the evaluation set includes a mix of head and tail queries, multiple verticals including retail, grocery, and convenience, diverse product categories, a non-trivial portion of misspellings and a very small set of multilingual queries.
Notably, around $30\%$ of unique queries in the evaluation set are unseen during training.
For each query, we sample items from both search logs and catalog to form $(q, d)$ pairs.
The proportion of irrelevant pairs per query is at least $65\%$, matching the production distribution.
For graded relevance, we map the three labels $2$, $1$, and $0$ to gains of $1$, $0.5$, and $0$ respectively.

Table \ref{tab:offline-main} reports relative performance of each model over the pre-trained backbone multilingual-E5-small.
In particular, pretrained models vary considerably on our domain-specific evaluation set regardless of model size, and our fine-tuned 0.1B model with 256-dimensional embeddings outperforms pretrained models up to 6× larger with 4× higher dimensionality, demonstrating the importance and effectiveness of domain-specific post-training over model scaling in this setting.

In addition to recall and precision, we also report NDCG to present the comprehensive evaluation of the incremental gain obtained by \emph{Mine-and-Refine} compared to the pre-trained baselines.
    
\subsubsection{Side-by-Side: Production Replica} 
We further evaluate end-to-end search relevance using Side-by-Side, an offline A/B framework: $\mytexttilde13$K queries sampled by annual engagement distribution are searched under a production replica (control) and the same system with the proposed embedding retriever swapped in (treatment), and NDCG@$10$ is compared. Specifically, the relevance labels in this practice are not generated by the aforementioned fine-tuned LLM labeler; instead, they are annotated by qualified experts to provide independent relevance quality evaluation that respects both business guidelines and true user intents.

Table \ref{tab:online-ab} reports a comparison of end-to-end performance across the lexical and hybrid baselines, and hybrid search retrieval with the proposed model, which shows incremental gains not only in retrieval but also in the quality of the downstream ranking results displayed to users.

\subsection{Online A/B Experiments}
\label{sec:online-ab}
We validate offline gains via two sequential online A/B tests in production search: (i) lexical baseline vs. hybrid baseline, followed by (ii) hybrid baseline vs. hybrid setup with the proposed model.
Both experiments were run for one month exposed to $90\%$ of the users with a $50\%/50\%$ traffic split. 
The treatment in each replaced only the embedding model, while the lexical retrieval, downstream ranking architecture, and business logic remained unchanged. 
In Table \ref{tab:online-ab}, we show the relative lifts in add-to-cart rate (ATCR), conversion rate (CVR), and the gross order value (GOV) for each variant against its respective A/B control, along with the associated p-values in parentheses.
Statistical significance is assessed using standard two-sample tests with a predefined threshold ($p<0.05$). 

\begin{table}[t]
    \centering
    \caption{Side by Side and Online A/B test results: relative improvement (\%) of Mine and Refine over the baseline system}
    \label{tab:online-ab}
    {\small 
    \begin{tabular}{l|cccc}
        \toprule
        System & \makecell[l]{Relevance \\ NDCG@$10$} & ATCR & CVR & GOV \\
        \midrule
        Lexical baseline & - & - & - & - \\ 
        Hybrid baseline & 1.66\% & \makecell{0.23\% \\ (0.00)} & \makecell{0.10\% \\ (0.01)} & \makecell{0.17\% \\ (0.04)}  \\   
        \makecell[l]{\textbf{\small Hybrid w/}\\ \textbf{\small Mine and Refine}} & \textbf{1.30\%} & \makecell{\textbf{1.04\%} \\ \textbf{(0.00)}} & \makecell{\textbf{1.27\%} \\ \textbf{(0.02)}} & \makecell{\textbf{0.44\%} \\ \textbf{(0.03)}} \\ 
        \bottomrule
    \end{tabular}%
    }
\noindent
\begin{minipage}{\linewidth}
\vspace{0.2cm}
\small
* Each variant's relative lift is reported against its immediately preceding control: hybrid baseline vs. lexical baseline, and Hybrid w/ Mine and Refine vs. hybrid baseline.
\end{minipage}
\end{table}

\begin{table}[htbp]
\centering
\caption{Ablation Studies on Model Architecture and Item Category Enrichment} 
\label{tab:structure-catalog-eval}
\setlength{\tabcolsep}{3pt}
{\small 
\begin{tabular}{ll|ccc|ccc}
\toprule
& & \multicolumn{3}{c|}{Recall@K}
& \multicolumn{3}{c}{Precision@K} \\
\cmidrule(lr){3-5}
\cmidrule(lr){6-8}
(Structure, Loss)
& Cat. 
& $10$ & $50$ & $100$
& $10$ & $50$ & $100$ \\
\midrule
(A, TL) & - & - & - & - & - & - & - \\
(A, CL) & - & -11.7\% & -11.0\% & -10.3\% & -11.4\% & -10.7\% & -10.3\% \\
(A, SC) & 4 & 1.1\% & 2.4\% & 2.5\% & 0.8\% & 1.9\% & 2.0\% \\
(A, SC) & - & 2.6\% & 2.8\% & 2.3\% & 2.4\% & 2.5\% & 2.0\% \\
(S, SC) & 4 & 6.5\% & 5.5\% & 4.7\% & 6.3\% & 5.5\% & 4.7\% \\
(S, SC) & - & 8.3\% & 5.8\% & 4.3\% & 8.2\% & 6.1\% & 4.6\% \\
\textbf{(S, SC)} & \textbf{2} & \textbf{8.7\%} & \textbf{6.2\%} & \textbf{6.0\%} & \textbf{9.1\%} & \textbf{6.7\%} & \textbf{4.5\%} \\
(S, SC) + $\text{(A, TL)}^{H\ddagger}$ & 2 & -30.4\% & -31.4\% & -31.4\% & -29.8\% & -29.0\% & -28.4\% \\
(S, SC) + (A, TL) & 2 & 8.8\% & 6.5\% & 7.5\% & 8.9\% & 7.7\% & 4.8\% \\
(S, SC) + (A, CL) & 2 & 11.6\% & 10.6\% & 9.5\% & 11.6\% & 11.2\% & 10.5\% \\
\textbf{(S, SC) + (S, CL)} & \textbf{2} & \textbf{12.3\%} & \textbf{11.0\%} & \textbf{9.8\%} & \textbf{12.3\%} & \textbf{11.5\%} & \textbf{10.6\%} \\
\bottomrule
\end{tabular}
}
\noindent
\begin{minipage}{\linewidth}
\small
\vspace{0.2cm}
\begin{itemize}[leftmargin=*, nosep]


\item[†] Notation: A = Asymmetric, S = Siamese; CL = Circle Loss, SC = SupCon Loss, TL = Triplet Loss.

\item[‡] Variant: $(\text{S, SC}) + \text{(A, TL)}^{H}$ has the same setup as $(\text{S, SC}) + \text{(A, TL)}$, except that negatives from the original training data are excluded during curriculum training.

\item[§] Category: “Cat.” denotes the level of item category path used in training and inference.

\end{itemize}

\end{minipage}
\end{table}

\subsection{Ablation Studies} \label{sec:ablation}
\subsubsection{Model Architecture}
Regarding the model architecture, the Siamese structure consistently outperforms asymmetric bi-encoders in both training stages, as shown in Tables~\ref{tab:offline-main} and~\ref{tab:structure-catalog-eval}.
Although an asymmetric encoder with untied parameters may allow the model to better fine-tune for each expertise (query and item)~\cite{Xie22}, we observe that keeping the Siamese structure yields slightly better performance.

\subsubsection{Hierarchical Item Category Enrichment}
The empirical results presented in Table~\ref{tab:structure-catalog-eval} show that incorporating category information improves disambiguation.
The optimal choice of granularity may diverge across different product categories, e.g. two levels are sufficiently clean for "\textit{Household > Home Decor > Other Home Decor}" while four levels are more informative for "\textit{Lifestyle > Clothing > Tops > Shirts}". Overall, more fine-grained category hierarchies introduce noise without commensurate gains in this case.

\subsubsection{Spelling Augmentation of Training Queries}
Table \ref{tab:spell-eval} presents the impact of several strategies for improving model robustness to spelling perturbations.
Regularizing between clean queries and their spelling variants is expected to encourage embedding invariance, but in practice leads to overfitting to query-specific patterns and increased sensitivity to spelling noise. 
In-place substitution similarly yields no measurable benefit. We conjecture that spelling variants presented without their clean counterparts act primarily as noise, as they lack well-formed semantic structure for the model to learn from.
In contrast, additive augmentation, as described in Section~\ref{sec:data}, substantially improves model quality without degrading clean-query performance.

\subsubsection{Synthetic Query Enrichment}
We keep the proportion of synthetic queries generated from item attributes relatively low, similar to  \cite{Xie22}. Such queries tend to exhibit higher-than-average lexical similarity to positive items, and over-representing them would bias the model toward lexical matching.
We observe that randomly injecting synthetic queries at a low ratio improves overall model performance on standard metrics but degrades separability. 
However, when synthetic queries are selectively added only for items lacking positive query–item pairs, and combined with an expanded item pool, the model achieves both improved performance and enhanced separability as is shown in Table~\ref{tab:syn-query}.

\begin{table}[htbp]
\centering
\caption{Ablation Study on Spelling Augmentation} 
\label{tab:spell-eval}
\setlength{\tabcolsep}{2pt}
{\small
\begin{tabular}{l|ccc|ccc}
\toprule
& \multicolumn{3}{c|}{Recall@K}
& \multicolumn{3}{c}{Precision@K} \\
\cmidrule(lr){2-4}
\cmidrule(lr){5-7}
Spelling Variations
& $K=10$ & $K=50$ & $K=100$
& $K=10$ & $K=50$ & $K=100$ \\
\midrule
{\small None} & -- & -- & -- & -- & -- & -- \\
{\small Regularization} & -2.3\% & -1.6\% & -1.5\% & -2.3\% & -1.7\% & -1.5\% \\
{\small In-place Substitution} & -7.0\% & -5.9\% & -5.3\% & -6.9\% & -5.9\% & -5.5\% \\
\textbf{{\small Additive Augmentation}} & \textbf{3.1\%} & \textbf{1.3\%} & \textbf{0.4\%} & \textbf{3.2\%} & \textbf{1.8\%} & \textbf{0.9\%} \\
\bottomrule
\end{tabular}
}
\end{table}

\begin{table}[htbp]
\centering
\caption{Ablation Study on Synthetic Query Enrichment} 
\label{tab:syn-query}
\setlength{\tabcolsep}{3pt}
{\small
\begin{tabular}{l|ccc|ccc|cc}
\toprule
& \multicolumn{3}{c|}{Recall@K}
& \multicolumn{3}{c|}{Precision@K}
& \multicolumn{2}{c}{Margin} \\
\cmidrule(lr){2-4}
\cmidrule(lr){5-7}
\cmidrule(lr){8-9}
Synthetic Queries
& 10 & 50 & 100 & 10 & 50 & 100 & Avg* & Worst* \\
\midrule
None 
& -- & -- & -- 
& -- & -- & -- 
& -- & -- \\

Random
& 6\% & 3\% & 2\%
& 6\% & 3\% & 3\%
& -13\% & -38\% \\

\textbf{Pos-missing + Expanded}
& \textbf{14\%} & \textbf{13\%} & \textbf{12\%}
& \textbf{14\%} & \textbf{13\%} & \textbf{12\%}
& \textbf{34\%} & \textbf{-4\%} \\
\bottomrule
\end{tabular}
}
\noindent
\begin{minipage}{\linewidth}
\small

\vspace{0.2cm}
* Margin metrics are computed on query--item pairs with high lexical similarity (token overlap $\geq 70\%$) and defined as:

Average Query level Margin:
$\mathbb{E}_{q}\!\left(\mathbb{E}_{d^{(2)}}\langle q,d^{(2)}\rangle - \mathbb{E}_{d^{(0)}}\langle q,d^{(0)}\rangle\right)$;\\
Worst case Margin:
$\mathbb{E}_{q}\!\left(\min_{d^{(2)}}\langle q,d^{(2)}\rangle - \max_{d^{(0)}}\langle q,d^{(0)}\rangle\right)$ where $d^{(2)}$ and $d^{(0)}$ are defined in equation \eqref{eq:tuples}.
\end{minipage}
\end{table}

\section{Training Details} \label{sec:parameters}
The Stage 1 training dataset are constructed from $\mytexttilde260$M query--item pairs with LLM-derived guideline-aligned labels; Stage 2 augments this with mined hard samples.

Training is performed on NVIDIA A10 GPUs using PyTorch~\cite{torch2019}.
The models are trained using the Adam~\cite{adam2015} optimizer with a base learning rate of $10^{-5}$ and layer-wise learning rate decay (factor=$0.9$).
The batch size is set to be $64$, where each instance is a query--item group in one of the formats defined in equation \eqref{eq:tuples}.

Regarding the hyper-parameters in the circle loss \eqref{eq:circle}, we set $\Delta_{2, p} = 0.75,\,\Delta_{1, n} = 0.6,\,\Delta_{1, p} = 0.4, \Delta_{0, n} = 0.25$ via grid search, with $O_{k, \cdot}$ chosen to be symmetric around optimal similarity of each level (i.e. $1$, $0.5$, $0$ for relevance labels $2$, $1$, $0$ respectively): $O_{2, p} = 1.25,\,O_{1, n} = 0.4,\,O_{1, p} = 0.6, O_{0, n} = - 0.25$.
As circle loss is found to be robust on the scale factor $\gamma$~\cite{Sun20}, we take a moderate value of $64$. 

\section{Conclusion}
In this paper, we introduce several mutually reinforcing techniques for training data construction and model optimization that advance small-scale retrieval embedding models in the domain-specific context of DoorDash. 
In particular, the \emph{Mine and Refine} strategy, LLM-supervised offline hard sample mining combined with a multi-level extension of circle loss, accounts for the majority of the performance gains, jointly addressing graded relevance under business relevance guidelines and similarity score separability critical for stable downstream score fusion and ranking. 
Beyond the core techniques, our experiments also surface practical lessons in broad applications: additive spelling augmentation outperforms invariance-based regularization, and selective synthetic query enrichment improves both retrieval metrics and score separability where random injection trades them off.
Following successful A/B experiments and relevance evaluation, the proposed embedding model has been deployed in production search across all non-restaurant verticals at DoorDash, including retail, grocery, and convenience.



\newpage
\section{GenAI Usage Disclosure}
This work utilizes generative AI tools in several aspects.
GPT-4o mini~\cite{OpenAI24} and OpenAI o3~\cite{OpenAI25} were used to audit and generate training labels, as described in Section \ref{sec:label}.
The model training code was developed by the authors. Claude Code with Claude Sonnet 4.5~\cite{sonnet4} was used for code review and minor corrections.
Claude Opus 4~\cite{opus4} and GPT-5.5~\cite{gpt5} were used to refine and polish initial drafts of Sections \ref{sec:intro} and \ref{sec:related-work}.
All other sections of the manuscript, including the training methodology, evaluation and experiments, ablation analyses, figures, tables, and the underlying raw data, were produced without the use of generative AI.
The authors reviewed and edited all AI-assisted content and take full responsibility for the work.

\bibliographystyle{ACM-Reference-Format}
\bibliography{reference}


\end{document}